\documentclass[pra,aps,groupedaddress,twocolumn,floatfix,nofootinbib]{revtex4}
\usepackage{graphicx}

\newcommand{\beq}{\begin{equation}}
\newcommand{\eeq}{\end{equation}}
\newcommand{\beqa}{\begin{eqnarray}}
\newcommand{\eeqa}{\end{eqnarray}}

\newcommand{\ket}[1]{\mbox{$ | #1 \rangle $}}

\def\half{\frac{1}{2}}
\def\opone{\leavevmode\hbox{\small1\normalsize\kern-.33em1}}

\begin{document}

\title{Optical Communication Without Photons}
\author{Nicolas Gisin}

\date{\small \today}
\maketitle

%\section{Counterfactual Quantum Communication}
%=======================
In a recent PRL Salih et al. \cite{Salih} presented a quantum communication protocol that allows one to communicate without any particle carrying the information from the sender to the receiver. At first sight, this is quite intriguing, as there is no telepathy, isn't it? Their setup involves a large number of nested interferometers and their claim is mathematically correct in the limit of infinitely many interferometers. However, the essential point I like to make can already be seen with only two nested interferometer, see Fig.1.

Alice, on the left, starts by sending a single photon into the left input port of her interferometer, the two other input ports have zero photons. Hence, the initial state is $\ket{1,0,0}$. After the first beam splitter the state reads $\cos(\theta_1)\ket{1,0,0}+i \sin(\theta_1)\ket{0,1,0}$, where $\theta_1$ characterizes the coupling ration in this first beam splitter (BS1). The second beam splitter (BS2) is a 50-50\% one, after this second beam splitter one has:
\beq
\cos(\theta_1)\ket{1,0,0}+i\frac{\sin(\theta_1)}{\sqrt{2}}\ket{0,1,0}-\frac{\sin(\theta_1)}{\sqrt{2}}\ket{0,0,1}
\eeq

Bob, on the right hand side, is the emitter (here the communication goes from right to left, as in arabic writing). In order to send the bit value $b=0$, he blocks his optical line, while to send $b=1$ he does nothing. For the clarity of exposition we add a central player, named Charlie, as indicated in Fig. 1. Charlie corresponds to the "Inner cycle" in figure 1 of \cite{Salih}. Let us compute the evolution in both cases, $b=0$ and $b=1$ (the third beam-splitter (BS3) in again 50-50\%, $\theta_2$ parameterizes the last beam splitter (BS4) coupling ratio and the third mode is ignored as it doesn't reach Alice's receivers, hence the states below are not normalized).

If $b=0$, the final state reads:
\beqa
\big(\cos(\theta_1)\cos(\theta_2)-\half\sin(\theta_1)\sin(\theta_2)\big)\ket{1,0} +\nonumber\\
i\big(\cos(\theta_1)\sin(\theta_2)+\half\sin(\theta_1)\cos(\theta_2)\big)\ket{0,1}
\eeqa

If $b=1$, the final state reads:
\beq
\cos(\theta_1)\cos(\theta_2)\ket{1,0} + i\cos(\theta_1)\sin(\theta_2)\ket{0,1}
\eeq

\begin{figure}
\centering
\includegraphics[width=1.3\columnwidth]{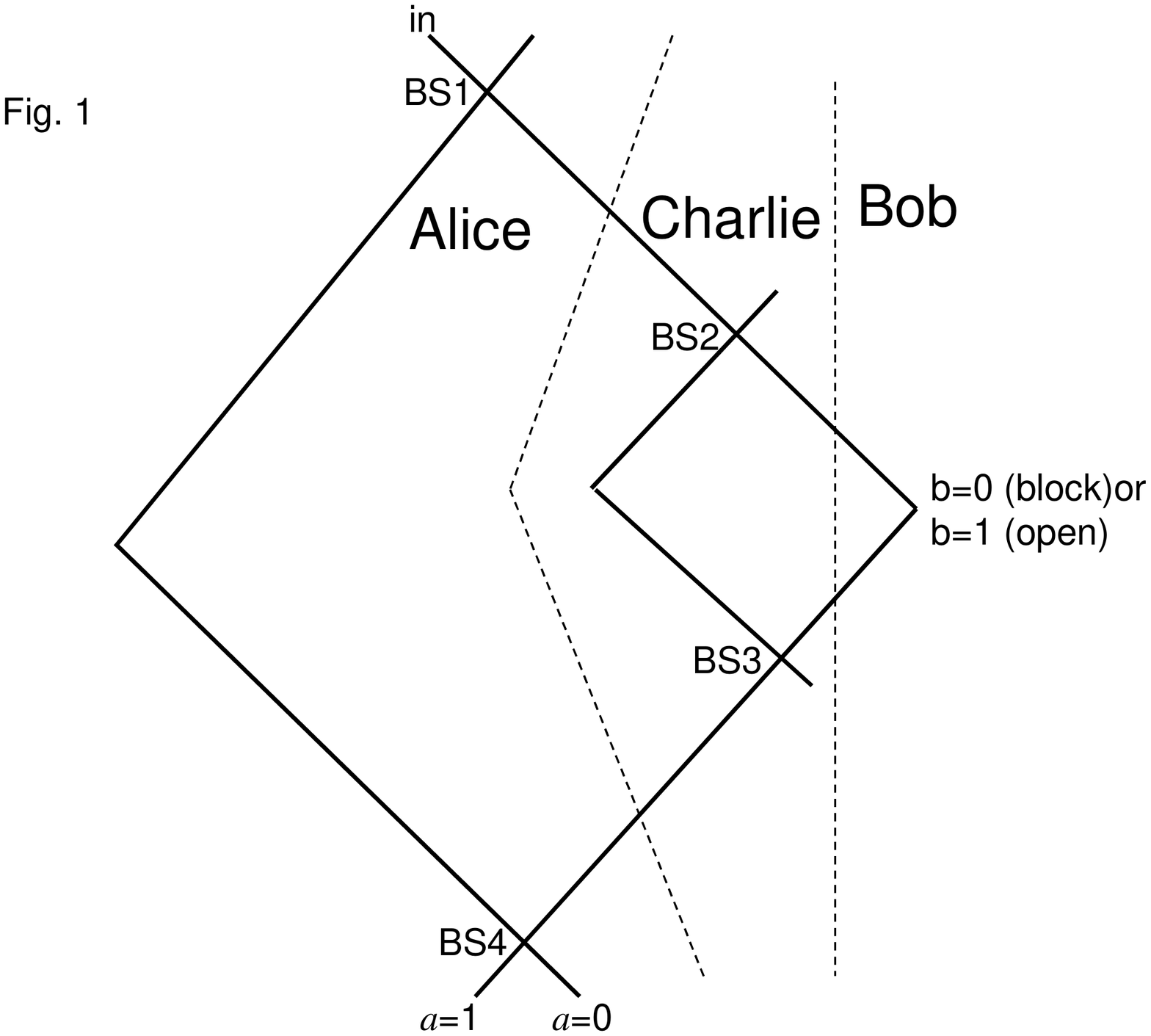}
\end{figure}

It is not difficult to find values of the coupling parameters $\theta_1$ and $\theta_2$ such that the probability that Alice's and Bob's bits are identical is larger than one half: $P(a=0|b=0)=P(a=1|b=1)>\half$ (e.g. $\theta_1=0.25$ and $\theta_2$ such that $\cos(\theta_2)^2=\frac{4\cos(\theta_1)^2}{\sin(\theta_1)^2-4\cos(\theta_1)\sin(\theta_1)+8\cos(\theta_1)^2}$). In such a case there is communication from Bob to Alice, though the amount of communication is less than one bit (Alice's detector may not register any photon or her result may be wrong). In \cite{Salih} it is shown that in the limit of infinitely many nested interferometers these limitations can be avoided such that Bob can communicate an entire bit in each run. Note though that, in the limit, each run takes infinite time; anyway, the amount of communication can be increased arbitrarily close to one bit.

Now, the surprising point made by Salih et al. is that when $b=0$, Bob blocks his optical line and thus no photon propagated from Bob to Charlie, while when $b=1$ the interference on the third beam-splitter is such that no photon propagates from Charlie to Alice. Consequently, there is communication from Bob to Alice without the photon ever propagating from Bob, the emitter, to Alice, the receiver!

Should one be shocked?

A first observation is that this the above scheme only involves linear optics (besides blocking Bob's optical mode). Hence, instead of one single-photon, Alice may as well send into her left input port a bright classical light pulse. Each photon of this classical pulse follows the same evolution as computed above. Hence, Alice can receive a full bit of information from Bob by merely using two classical (linear) detectors and reading her bit off the detector registering the largest signal. We have now a relatively simple communication channel from Bob to Alice allowing them to communicate one bit per use of the channel. Moreover, when Bob blocks, i.e. $b=0$, then no photon propagates from Bob to Charlie and when Bob leave his optical mode open, i.e. $b=1$, then no photon propagates from Charlie to Alice.

But now everything is classical (and could equally be realized with water waves). So what is the catch?

Consider the following communication protocol sketched in Fig. 2. First, Alice sends two (classical) billiard balls to Charlie, a red and a blue one. Next, Charlie sends the blue ball to Bob. If Bob likes to transmit $b=0$, he keeps the blue ball; if he chooses $b=1$, he sends the blue ball back to Charlie. Thirdly, if Charlie doesn't receive the blue ball, he sends back the red ball to Alice; but if he receives the blue ball, then he keeps the red one. In this way, if Alice receives no ball, she knows that $b=1$, while if she receives the red ball, she knows $b=0$. This is a perfect communication channel, not difficult to implement, despite the fact that there is no ball that goes from Bob, the emitter, to Alice, the receiver.

The above protocol is actually quite close to standard classical optical communication. Indeed, in the simplest version of classical optical communication, the emitter sends either a bright light pulse ($b=1$) or "nothing", i.e. an empty pulse, ($b=0$). Hence, in this traditional communication protocol, in half the cases "nothing" carries the communicated bit. This protocol can be adapted to mimic our protocol above. First, let the emitter, Bob, encode his bits with full and empty pulses that he sends to Charlie. Next, Charlie reads the bits, flips them all and encodes the flipped bit in empty and full pulses that he sends to Alice. Alice decodes the pulses with the convention $a=0$ whenever the pulse is full and $a=1$ whenever the pulse is empty. In this way Alice gets all of Bob's bits correctly, with unit probability, despite that, for each bit, no photon travels all the way from Bob to Alice.

The lesson from this is not new: vacuum is not nothing. To communicate the emitter and receiver must be connected by a channel. In the case of optical communication, whether quantum or classical, this channel is an optical mode. The information carrier are quantum or classical states of this optical mode, including possibly the vacuum (empty pulse).

\small
\section*{Acknowledgment}
This note profited from stimulating exchanges with Tomer Barnea, Yeong-Cherng Liang, Gilles Puetz, Pavel Sekatski and Nikola Vona.
This work has been supported by the ERC-AG grant QORE and by the Swiss NCCR-QSIT.

\end{document}